\documentclass[journal]{IEEEtran}
\ifCLASSINFOpdf \else \fi

\usepackage{graphicx}
\usepackage[cmex10]{amsmath}
\usepackage{subfigure}
\usepackage{amssymb}

\usepackage[table]{xcolor}
\usepackage{tabularx}
\usepackage{multirow}
\usepackage{acro}[=v3] 
\DeclareAcronym{RIS}{
  short = RIS ,
  long  = reconfigurable intelligent surfaces ,
  tag = abbrev
}
\DeclareAcronym{6G}{
  short = 6G,
  long  = sixth generation ,
  tag = abbrev
}
\DeclareAcronym{5G}{
  short = 5G,
  long  = 5th generation ,
  tag = abbrev
}
\DeclareAcronym{RDM}{
  short =   RDM,
  long  = range-Doppler map ,
  tag = abbrev
}
\DeclareAcronym{ISAC}{
  short = ISAC ,
  long  = integrated sensing and communication,
  tag = abbrev
}
\DeclareAcronym{SA-OFDM}{
  short = SA-OFDM ,
  long  = subcarrier aliasing OFDM,
  tag = abbrev
}
\DeclareAcronym{EM}{
  short = EM ,
  long  = electromagnetic,
  tag = abbrev
}
\DeclareAcronym{BS}{
  short = BS,
  long  = base station ,
  tag = abbrev
}
\DeclareAcronym{UE}{
  short = UE ,
  long  = user equipment ,
  tag = abbrev
}
\DeclareAcronym{MISO}{
  short = MISO,
  long  = multiple-input single-output ,
  tag = abbrev
}
\DeclareAcronym{MMSE}{
  short = MMSE ,
  long  = minimum mean squared error ,
  tag = abbrev
}
\DeclareAcronym{RMSE}{
  short = RMSE ,
  long  = root mean squared error ,
  tag = abbrev
}
\DeclareAcronym{DFT}{
  short = DFT,
  long  = discrete Fourier transform ,
  tag = abbrev
}
\DeclareAcronym{IDFT}{
  short = IDFT,
  long  = inverse discrete Fourier transform ,
  tag = abbrev
}
\DeclareAcronym{FFT}{
  short = FFT,
  long  = fast Fourier transform ,
  tag = abbrev
}
\DeclareAcronym{ISFFT}{
  short = ISFFT,
  long  = inverse symplectic finite Fourier transform ,
  tag = abbrev
}
\DeclareAcronym{IFFT}{
  short = IFFT,
  long  = inverse fast Fourier transform ,
  tag = abbrev
}
\DeclareAcronym{SFFT}{
  short = SFFT,
  long  = symplectic finite Fourier transform ,
  tag = abbrev
}
\DeclareAcronym{THz}{
  short = THz,
  long  = Terahertz ,
  tag = abbrev
}
\DeclareAcronym{MPC}{
  short = MPC,
  long  = multipath component ,
  tag = abbrev
}
\DeclareAcronym{IoT}{
  short = IoT,
  long  = internet of things ,
  tag = abbrev
}
\DeclareAcronym{V2X}{
  short = V2X,
  long  = vehicle to everything ,
  tag = abbrev
}
\DeclareAcronym{V2V}{
  short = V2V,
  long  = vehicle to vehicle ,
  tag = abbrev
}
\DeclareAcronym{MSE}{
  short = MSE,
  long  = mean square error ,
  tag = abbrev
}
\DeclareAcronym{CSI}{
  short = CSI ,
  long  = channel state information ,
  tag = abbrev
}
\DeclareAcronym{MIMO}{
  short = MIMO,
  long  = multiple-input multiple-output ,
  tag = abbrev
}
\DeclareAcronym{UPA}{
  short = UPA,
  long  = uniform planner array ,
  tag = abbrev
}
\DeclareAcronym{RF}{
  short = RF,
  long  = radio-frequency ,
  tag = abbrev
}
\DeclareAcronym{mmWave}{
  short = mmWave,
  long  = millimeter-wave ,
  tag = abbrev
}
\DeclareAcronym{AoA}{
  short = AoA ,
  long  = angle of arrival ,
  tag = abbrev
}
\DeclareAcronym{AoD}{
  short = AoD,
  long  = angle of departure ,
  tag = abbrev
}
\DeclareAcronym{EKF}{
  short = EKF,
  long  = extended Kalman filter ,
  tag = abbrev
}
\DeclareAcronym{LMS}{
  short = LMS,
  long  = least mean square ,
  tag = abbrev
}
\DeclareAcronym{BiLMS}{
  short = BiLMS,
  long  = bi-directional LMS ,
  tag = abbrev
}
\DeclareAcronym{SNR}{
  short = SNR,
  long  = signal-to-noise ratio ,
  tag = abbrev
}
\DeclareAcronym{LoS}{
  short = LoS,
  long  = line-of-sight ,
  tag = abbrev
}
\DeclareAcronym{NLoS}{
  short = NLoS,
  long  = non line-of-sight ,
  tag = abbrev
}

\DeclareAcronym{NMSE}{
  short = NMSE,
  long  = normalized mean square error ,
  tag = abbrev
}
\DeclareAcronym{PAPR}{
  short = PAPR,
  long  = peak to average power ratio ,
  tag = abbrev
}
\DeclareAcronym{PDP}{
  short = PDP,
  long  = power delay profile ,
  tag = abbrev
}
\DeclareAcronym{ISI}{
  short = ISI,
  long  = inter-symbol interference ,
  tag = abbrev
}
\DeclareAcronym{ICI}{
  short = ICI,
  long  = inter-carrier interference ,
  tag = abbrev
}
\DeclareAcronym{SDR}{
  short = SDR,
  long  = semidefinite relaxation ,
  tag = abbrev
}
\DeclareAcronym{QoS}{
  short = QoS,
  long  = quality of service ,
  tag = abbrev
}
\DeclareAcronym{NOMA}{
  short = NOMA,
  long  = non-orthogonal multiple access ,
  tag = abbrev
}
\DeclareAcronym{OMA}{
  short = OMA,
  long  = orthogonal multiple access ,
  tag = abbrev
}
\DeclareAcronym{NU}{
  short = NU,
  long  = near user ,
  tag = abbrev
}
\DeclareAcronym{CIR}{
  short = CIR,
  long  = channel impulse response ,
  tag = abbrev
}
\DeclareAcronym{FU}{
  short = FU,
  long  = far user ,
  tag = abbrev
}
\DeclareAcronym{CP}{
  short = CP,
  long  = cyclic prefix ,
  tag = abbrev
}
\DeclareAcronym{ZP}{
  short = ZP,
  long  = zero padding ,
  tag = abbrev
}
\DeclareAcronym{ZS}{
  short = ZS,
  long  = zero suffix ,
  tag = abbrev
}
\DeclareAcronym{ZF}{
  short = ZF,
  long  = zero forcing ,
  tag = abbrev
}
\DeclareAcronym{RCP}{
  short = RCP,
  long  = reduced-CP ,
  tag = abbrev
}
\DeclareAcronym{FZS}{
  short = FZS,
  long  = full-ZS ,
  tag = abbrev
}
\DeclareAcronym{RZP}{
  short = RZP,
  long  = reduced-zero padded ,
  tag = abbrev
}
\DeclareAcronym{FCP}{
  short = FCP,
  long  = full-CP ,
  tag = abbrev
}
\DeclareAcronym{RFCP}{
  short = RFCP,
  long  = reduced-full-CP ,
  tag = abbrev
}
\DeclareAcronym{BER}{
  short = BER,
  long  = bit error rate ,
  tag = abbrev
}
\DeclareAcronym{SIC}{
  short = SIC,
  long  = successive interference cancellation ,
  tag = abbrev
}
\DeclareAcronym{PLS}{
  short = PLS,
  long  = physical layer security ,
  tag = abbrev
}
\DeclareAcronym{MRT}{
  short = MRT,
  long  = maximum ratio transmission ,
  tag = abbrev
}
\DeclareAcronym{AWGN}{
  short = AWGN,
  long  = additive white Gaussian noise,
  tag = abbrev
}
\DeclareAcronym{SINR}{
  short = SINR,
  long  = signal-to-interference-plus-noise ratio ,
  tag = abbrev
}
\DeclareAcronym{BPSK}{
  short = BPSK,
  long  = binary phase shift keying ,
  tag = abbrev
}
\DeclareAcronym{QPSK}{
  short = QPSK,
  long  = quadrature phase shift keying ,
  tag = abbrev
}
\DeclareAcronym{SVD}{
  short = SVD,
  long  = singular value decomposition ,
  tag = abbrev
}
\DeclareAcronym{EVD}{
  short = EVD,
  long  = eigenvalue decomposition ,
  tag = abbrev
}
\DeclareAcronym{PDF}{
  short = PDF,
  long  = probability density function ,
  tag = abbrev
}
\DeclareAcronym{SER}{
  short = SER,
  long  = symbol error rate ,
  tag = abbrev
}
\DeclareAcronym{MGF}{
  short = MGF,
  long  = moment generating function ,
  tag = abbrev
}
\DeclareAcronym{2D}{
  short = 2D,
  long  = two-dimensional ,
  tag = abbrev
}
\DeclareAcronym{3D}{
  short = 3D,
  long  = three-dimensional ,
  tag = abbrev
}
\DeclareAcronym{CLT}{
  short = CLT,
  long  = central limit theorem ,
  tag = abbrev
}
\DeclareAcronym{QAM}{
  short = QAM,
  long  = quadrature amplitude modulation ,
  tag = abbrev
}
\DeclareAcronym{SISO}{
  short = SISO,
  long  = single-input single-output ,
  tag = abbrev
}
\DeclareAcronym{CE}{
  short = CE,
  long  = channel estimation ,
  tag = abbrev
}
\DeclareAcronym{KG}{
  short = $K_G$,
  long  = generalized-K ,
  tag = abbrev
}
\DeclareAcronym{LSKRF}{
  short = LSKRF,
  long  = least squares Khatri-Rao factorization ,
  tag = abbrev
}
\DeclareAcronym{FMCW}{
  short = FMCW,
  long  = frequency modulated continuous wave ,
  tag = abbrev
}
\DeclareAcronym{FSK}{
  short = FSK,
  long  = frequency shift keying ,
  tag = abbrev
}
\DeclareAcronym{JSAC}{
  short = JSAC,
  long  = joint sensing and communication ,
  tag = abbrev
}
\DeclareAcronym{OTFS}{
  short = OTFS,
  long  = orthogonal time-frequency space ,
  tag = abbrev
}
\DeclareAcronym{OTSM}{
  short = OTSM,
  long  = orthogonal time sequency multiplexing ,
  tag = abbrev
}
\DeclareAcronym{ADAS}{
  short = ADAS,
  long  = Advanced driver assistant system ,
  tag = abbrev
}
\DeclareAcronym{MP}{
  short = MP,
  long  = message passing ,
  tag = abbrev
}
\DeclareAcronym{ML}{
  short = ML,
  long  = maximum likelihood ,
  tag = abbrev
}
\DeclareAcronym{OFDM}{
  short = OFDM,
  long  = orthogonal frequency division multiplexing ,
  tag = abbrev
}
\DeclareAcronym{PSD}{
  short = PSD,
  long  = power spectral density,
  tag = abbrev
}
\DeclareAcronym{SE}{
  short = SE,
  long  = spectral efficiency,
  tag = abbrev
}
\DeclareAcronym{AFDM}{
  short = AFDM,
  long  = affine frequency division multiplexing ,
  tag = abbrev
}
\DeclareAcronym{OCDM}{
  short = OCDM,
  long  = orthogonal chirp division multiplexing ,
  tag = abbrev
}
\DeclareAcronym{ODDM}{
  short = ODDM,
  long  = orthogonal delay-Doppler division multiplexing,
  tag = abbrev
}
\DeclareAcronym{LTV}{
  short = LTV,
  long  = linear time-varying,
  tag = abbrev
}
\DeclareAcronym{LTI}{
  short = LTI,
  long  = linear time-invarying,
  tag = abbrev
}
\DeclareAcronym{NTN}{
  short = NTN,
  long  = non-terrestrial network,
  tag = abbrev
}

\DeclareAcronym{TN}{
  short = TN,
  long  = terrestrial network,
  tag = abbrev
}

\DeclareAcronym{AI}{
  short = AI,
  long  = artificial intelligence,
  tag = abbrev
}
\DeclareAcronym{LEO}{
  short = LEO,
  long  = low Earth orbit,
  tag = abbrev
}
\DeclareAcronym{TF}{
  short = T-F,
  long  = time-frequency,
  tag = abbrev
}
\DeclareAcronym{DD}{
  short = DD,
  long  = delay-Doppler,
  tag = abbrev
}
\DeclareAcronym{AF}{
  short = AF,
  long  = ambiguity function,
  tag = abbrev
}
\DeclareAcronym{LoRa}{
  short = LoRa,
  long  = Long-Range,
  tag = abbrev
}
\DeclareAcronym{EVA}{
  short = EVA,
  long  = Extended Vehicular A,
  tag = abbrev
}
\DeclareAcronym{SMT}{
  short = SMT ,
  long  = sparse multi-tap,
  tag = abbrev
}
\DeclareAcronym{TDL}{
  short = TDL ,
  long  = tapped delay line,
  tag = abbrev
}

\DeclareAcronym{FDD}{
  short = FDD ,
  long  = frequency division duplexing,
  tag = abbrev
}
\DeclareAcronym{TDD}{
  short = TDD ,
  long  = time division duplexing,
  tag = abbrev
}

\DeclareAcronym{IBFD}{
  short = IBFD ,
  long  = in band full duplex,
  tag = abbrev
}
\DeclareAcronym{SBFD}{
  short = SBFD ,
  long  = sub-band full duplex,
  tag = abbrev
}
\DeclareAcronym{DSP}{
  short =DSP ,
  long  = digital signal processing,
  tag = abbrev
}
\DeclareAcronym{OOK}{
  short =OOK ,
  long  = ON-OFF keying,
  tag = abbrev
}
\DeclareAcronym{RCS}{
  short = RCS,
  long  = radar cross section,
  tag = abbrev
}
\DeclareAcronym{URLLC}{
  short = URLLC,
  long  = ultra reliable low latency communication,
  tag = abbrev
}
\usepackage{algorithmic}
\usepackage{makecell}
\usepackage{ragged2e}
\usepackage{array}
\usepackage{url}
\usepackage{cite}
\usepackage[compact]{titlesec}
\titlespacing{\section}{1.0pt}{*1.0}{*0}
\titlespacing{\subsection}{1.1pt}{*1.1}{*0}
\titlespacing{\subsubsection}{0.3pt}{*0}{*0}
\begin{document}
\bstctlcite{IEEEexample:BSTcontrol}
\title{A Unified Framework for Adaptive Waveform Processing in Next Generation Wireless Networks}
\author{
\IEEEauthorblockN{Abdelali Arous,~\IEEEmembership{Graduate~Student~Member,~IEEE}, Hamza Haif,~\IEEEmembership{Graduate~Student~Member,~IEEE}, \\ Arman Farhang,~\IEEEmembership{Senior Member,~IEEE} and H\"{u}seyin Arslan,}
\IEEEmembership{Fellow, IEEE}
\thanks{The authors A. Arous, H. Haif, and H. Arslan, are with the Department of Electrical and Electronics Engineering, Istanbul Medipol University, Istanbul, 34810, Turkey, (e-mail: abdelali.arous@std.medipol.edu.tr; hamza.haif@std.medipol.edu.tr;  huseyinarslan@medipol.edu.tr). Arman Farhang is with Department of Electronic and Electrical Engineering, Trinity College Dublin, Dublin, Ireland (e-mail: arman.farhang@tcd.ie)}
}


\maketitle

\begin{abstract}
The emergence of alternative multiplexing domains to the time-frequency domains, e.g., the delay-Doppler and chirp domains, offers a promising approach for addressing the challenges posed by complex propagation environments and next-generation applications. Unlike the time and frequency domains, these domains offer unique channel representations which provide additional degrees of freedom (DoF) for modeling, characterizing, and exploiting wireless channel features. This article provides a comprehensive analysis of channel characteristics, including delay, Doppler shifts, and channel coefficients across various domains, with an emphasis on their inter-domain relationships, shared characteristics, and domain-specific distinctions. We further evaluate the comparative advantages of each domain under specific channel conditions. Building on this analysis, we propose a generalized and adaptive transform domain framework that leverages the pre- and post-processing of the discrete Fourier transform (DFT) matrix, to enable dynamic transitions between various domains in response to the channel conditions and system requirements. Finally, several representative use cases are presented to demonstrate the applicability of the proposed cross-domain waveform processing framework in diverse scenarios, along with future directions and challenges. 
\end{abstract}
\IEEEpeerreviewmaketitle
\vspace{-1mm}
\section{Introduction}
\IEEEPARstart{U}{nder} the International Mobile Telecommunications (IMT)-2030 framework, the \ac{6G} of wireless networks is anticipated to support advanced applications such as \ac{ISAC}, large-scale \ac{IoT} deployments, secure and high-mobility communications \cite{saad2019vision}. Achieving these ambitious objectives necessitates fundamental modifications at the physical layer, particularly through novel signal designs, prompting the exploration of alternative waveforms to the conventional \ac{OFDM}, such as \ac{DD} and chirp-based multicarrier modulations. However, a complete replacement of \ac{OFDM}, the cornerstone of fourth- and fifth-generation wireless networks (4G/5G), is constrained by the imperative of backward compatibility, as mandated by the 3rd generation partnership project (3GPP) standardization \cite{3gpp_tr_38_900}. Backward compatibility aspects include transceiver architecture, pilot structures, scheduling mechanisms, \ac{MIMO} capabilities and dynamic spectrum management. Notably, emerging waveforms like \ac{OTFS} and \ac{AFDM} can be interpreted as \ac{DFT}-based pre- and post-processing extensions of the OFDM demodulator, enabling seamless integration with current 3GPP compliant waveforms \cite{solaija2024orthogonal}. By multiplexing data in alternative logical domains (e.g., DD or Fresnel/affine), these waveforms offer distinct channel representations beyond what is observed in \ac{TF} domain, unlocking new capabilities under diverse channel conditions \cite{xiao2024rethinking}. Nonetheless, rather than functioning as standalone multiplexing schemes, can we synergistically exploit logical (DD/chirp) and physical (T-F) domains to develop a unified signal processing framework that is backward compatible with the 3GPP 4G and 5G radio standards?
\par DD domain waveforms, such as \ac{OTFS} and \ac{ODDM}, employ a two-dimensional (2D) grid where wireless channels exhibit sparse and quasi-static structures \cite{deng2025unifying}. Conversely, chirp-based waveforms like \ac{OCDM} and \ac{AFDM} utilize one-dimensional (1D) chirp carriers that span the entire \ac{TF} plane. While \ac{OCDM} relies on fixed chirp slopes, \ac{AFDM} dynamically adjusts the chirp tilt to align with the channel's maximum Doppler spread \cite{AFDM_fractional}. These spreading mechanisms, whether 1D or 2D, provide resilience against doubly dispersive channels, achieve full \ac{TF} diversity, and enable \ac{ISAC} functionalities. Furthermore, these logical domains introduce an additional degree of freedom (DoF) for characterizing channel delays, Doppler shifts, and coefficients beyond the conventional \ac{TF} analysis. Specifically, they offer unique channel representations in terms of sparsity, \ac{MPC}s resolvability, and quasi-static behavior, especially in high-dynamic scenarios. For instance, the DD domain’s inherent alignment with physical reflectors facilitates efficient \ac{ISAC} capability, while chirp-based signals support low-complexity  receiver designs suitable for energy-constrained IoT devices. Consequently, these logical domains should be viewed as complements, not replacements, to traditional T-F techniques, as they offer refined channel modeling, performance prediction, and reliability improvements. 
The ability to employ these waveforms interchangeably allows for their integration alongside existing \ac{OFDM} systems, eliminating the need for a complete transition to a new waveform. This approach addresses the intrinsic limitations of \ac{OFDM} while enabling the performance gains demonstrated by emerging 6G waveform candidates.
\par This magazine presents a comprehensive analysis of channel representations across physical and logical domains under idealized and complex propagation environments, emphasizing delay-Doppler and channel coefficient dynamics and their impact on \ac{OTFS}, \ac{AFDM}, and \ac{OFDM} performance. We then propose a generalized, adaptive waveform processing framework that leverages domain-specific channel features for cross-domain data-multiplexing, \ac{CE}, equalization, and \ac{ISAC} integration. Finally, we propose practical use cases leveraging the proposed framework to enable different applications, followed by practical challenges and future directions.

\section{Channel Parameters Interpretation}
\begin{figure*}[t]
   \centering
   \includegraphics[scale=0.17]{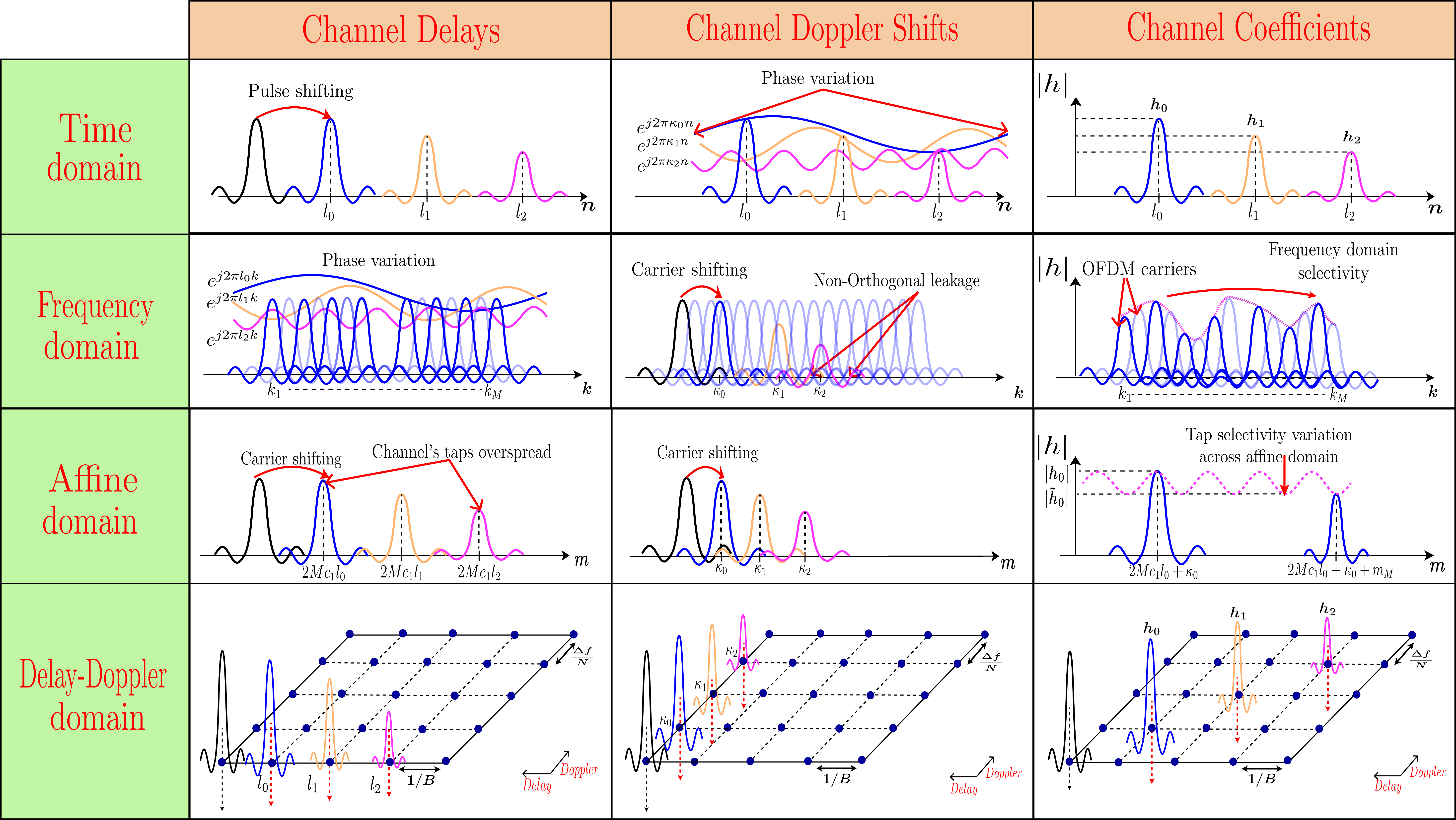}
   \caption{Behavior of channel coefficients, delays, and Doppler shifts across various waveform domains: the black pulse represents the transmitted pulse, whereas the colored pulses denote the received channel MPCs.}
   \label{fig: hhhh}
\end{figure*}

This section provides a comparative representation of the channel parameters in the aforementioned physical and logical domains. The analysis assumes a sparse and resolvable multipath propagation environment, where each \ac{MPC} indexed by $i$ is characterized with a channel coefficient $h_i$ and distinct normalized delay $l_i$, and normalized Doppler shift $\kappa_i$. The channel representation in the time, frequency, affine (or chirp) and \ac{DD} domains are illustrated in Fig. \ref{fig: hhhh}.
\subsection{Delay Shifts Representation}
In the time domain, \ac{MPC}s that induce temporal dispersion can be resolved individually depending on the signal bandwidth $B$, where the minimum resolvable delay is given by $\Delta_{\min} = \frac{1}{B}$. These delay components are conventionally modeled through the \ac{PDP} and its associated delay spread, which are critical for the design of \ac{CP} in multicarrier waveforms.  These delays manifest as cumulative phase rotations across subcarriers in the frequency domain. This property facilitates low-complexity single-tap equalization for \ac{OFDM} in frequency-selective channels. Alternatively, the representation of these \ac{MPC}s in other multiplexing domains is given by: 
\par \textit{\textbf{Chirp domain}}: The normalized delay  appears as a scaled shift in the Fresnel/affine domains, proportional to the design parameter $c_1$. While this representation maintains the delay shift property of the \ac{MPC}s as in the time domain, the scaling will convert an underspread channel with close-by MPCs into an overspread channel in the affine domain. This spread leads to extended pilot guard sizes, reducing the \ac{SE} proportionally. Moreover, these delay shifts induce a rapidly varying quadratic phase term in the affine domain, causing different data symbols to experience varying channel effects \cite{arous2024novel}.    
\par \textit{\textbf{DD domain}}: Similar to the time domain, the 2D orthogonal \ac{DD} plane permits the \ac{MPC}s to appear as distinct shifts along the delay axis in separate bins. Thereby, the channel has a sparse, structured and quasi-static representation providing symbol-level diversity and \ac{MPC}s resolvability. Additionally, the delay shifts cause a non-dominant linear phase variation across the DD domain. This phase variation is also a function of the adopted \ac{CP} configuration, whether it is full or reduced \ac{CP}.    
\subsection{Doppler Shifts Representation}
With the increasing prevalence of high-mobility devices in modern wireless networks, and the emergence of FR3 technology which covers bands up to 24.25 GHz, the Doppler impact becomes more pronounced and less tractable compared to FR1. Thus, accurately characterizing Doppler behavior across various signal domains is essential for the design of robust communication systems.
\par \textit{\textbf{Physical domains}}: Due to the \ac{TF} duality property inherent in the Fourier transform, Doppler shifts manifested as linear phase variations in the time domain appear as frequency shifts in the frequency domain. This reciprocal relationship to MPCs delays implies that Doppler-induced time selectivity destroys orthogonality in multicarrier waveforms such as \ac{OFDM}, resulting in \ac{ICI}. The resolvability of Doppler shifts in the frequency domain is proportional to the subcarrier spacing; hence, having multi-numerology options as in 5G improves the ability to distinguish between different Doppler components.   
\par \textit{\textbf{Logical domains}}: In the DD and chirp-based domains, the Doppler shift retains a shift-invariant structure, although with distinct behaviors. In the affine domain, Doppler shifts appear coupled with the scaled delay shifts. Same property is consistent in the Fresnel domain, without any delay scaling. This coupling leads to complex ICI patterns in the multiplexing domain, which necessitates joint estimation and mitigation of both delay and Doppler components. The resolvability of Doppler shifts in these domains is determined by the frequency spacing between chirp carriers of AFDM and OCDM. Conversely, within the DD domain, Doppler manifests as separated shifts along the Doppler axis, while simultaneously inducing a linear phase variation. In \ac{OTFS} modulation, this Doppler representation benefits from enhanced resolution, specifically, an $N$-fold increase compared to the frequency and chirp domains, owing to the spreading of data symbols over an extended temporal occupation, where $N$ denotes the number of Doppler bins. This makes the DD domain particularly favorable for high-mobility scenarios.

\subsection{Channel Coefficients}
\par \textit{\textbf{Physical domains}}:
In the time domain, the transmitted signals affected by the MPCs arrive at distinct delay instances, each associated with an individual channel coefficient whose magnitude typically follows Rayleigh or Rician fading distributions. However, in the frequency domain, the channel response at each carrier is a coherent sum of the contributions from all MPCs, each weighted by its complex gain and phase rotation due to delay shifts. This frequency-domain superposition renders the separability of individual channel coefficients impractical. 
\par \textit{\textbf{Logical domains}}: 
In the affine domain, each MPC results in a localized channel tap aligned with its corresponding delay and Doppler-induced shift. Nevertheless, due to the phase-wrapping and rapid variation across the multiplexing plane, the effective magnitude of the channel coefficients becomes highly varying. These variations are influenced not only by the delay-induced phase but also by the design parameters $c_1$ and $c_2$. Consequently, relying solely on the parametrized guard estimated coefficients for equalization without appropriate phase compensation leads to substantial \ac{BER} degradation. Alternatively, as the DD domain gives a direct mapping to the MPCs, it preserves the separated channel coefficients representation in the DD grid.

\section{Special Channel Effects}
This section addresses special channel behavior under the interaction of the channel effects, hardware and resource limitations, and transform domain properties.
\subsection{Fractional Spread}
\begin{figure*}[t]
   \centering
   \includegraphics[scale=0.17]{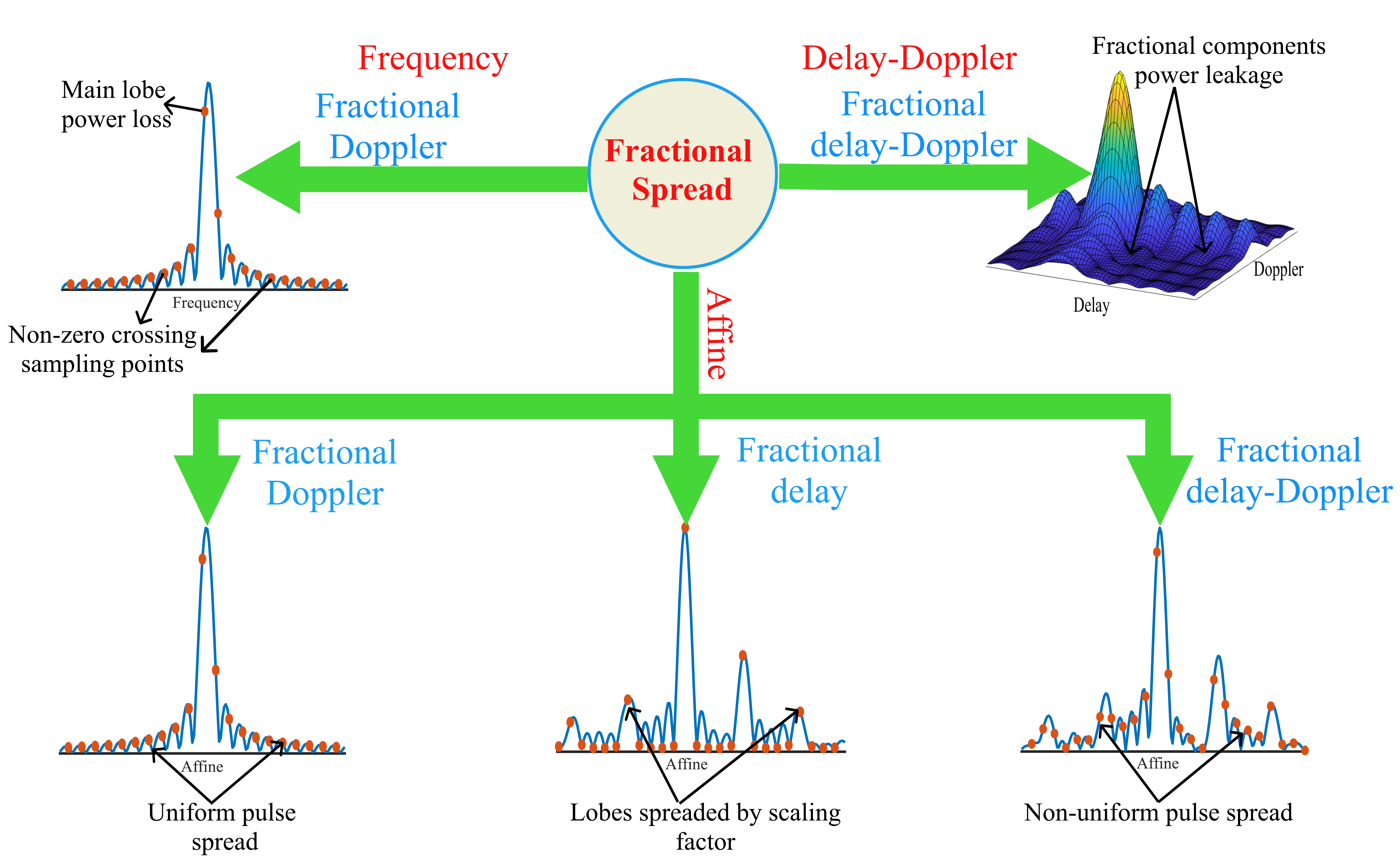}
   \caption{Channel's delay and Doppler fractional spread behavior in different multiplexing domains, under time domain rectangular pulse shaping.}
   \label{fig: hh}
\end{figure*}

The randomness of the spatial distribution and velocities of reflectors within the environment elevates the probability that channel delay and Doppler shifts manifest as fractional multiples of the sampling interval and frequency granularity of the transmitted signal. The impact of fractional spread is observed in the data multiplexing domain, with its severity contingent upon the specific waveform transform and its interaction with the \ac{TF} dispersion. For instance, due to the Fourier transform's property of mapping Doppler as frequency shifts, and delays as summed phase, \ac{OFDM} is primarily affected by fractional Doppler spread, leading to \ac{ICI}. Conversely, emerging waveforms that capture delays as shifts introduce an additional layer of fractional spread, resulting in a dual source of inter-data interference. This converts the sparse channel representation of these logical domains into a rich and overlapping effective channel MPCs, rendering \ac{CE} highly susceptible to errors. While \ac{DD}-based waveforms isolate fractional delay spread from fractional Doppler spread through their 2D grid orthogonality \cite{OTFS_fraction}, multicarrier chirp-based waveforms exhibit ambiguous fractional spread behavior due to the coupling of delay and Doppler and the quadratic rotation phase \cite{AFDM_fractional}. Moreover, the fractional spread source has an impact on the pulse shape behavior in the chirp domain, as depicted in Fig. \ref{fig: hh}, and explained below:  
\begin{itemize}
    \item Fractional Doppler: the pulse maintains the same shape shifted by the equivalent fractional component. 
    \item Fractional delay: the pulse fractional components will be spaced proportionally to the scaling factor of the transform coefficient $c_1$.
    \item Coupled fractional spread: both delay and Doppler fractional spreads create a non-uniform pulse shape, with both scaling and shifting.
\end{itemize}
Fractional spread mitigation is based on its source; fractional Doppler can be contained by employing larger subcarrier spacing and through time domain pulse shaping design, either through Nyquist pulses like raised cosine or non-Nyquist pulses such as Gaussian. Conversely, fractional delay spread necessitates spectrum shaping. Thus, the mitigation of both delay and Doppler spreads requires local and global windowing, which significantly increases system design complexity.
\subsection{Overspread Channel}
Doubly overspread channels, such as those encountered in underwater and \ac{V2X} channels, are characterized by simultaneously large delay spreads and Doppler spreads, with their product exceeding one. To mitigate the impact of Doppler spread in \ac{OFDM}, the subcarrier spacing is typically increased. However, this adjustment inherently reduces the symbol duration, where the hard-coded \ac{CP} set to match the maximum delay spread will be comparable to the symbol duration, causing significant \ac{SE} loss. For other waveforms, the capturing of the Doppler and delay overspreads requires further spreading in time along with wider bandwidths, resulting in significant normalized delay and Doppler shifts. This entails adopting larger pilot guard intervals to suppress inter-pilot-data interference, further reducing the effective \ac{SE}. Additionally, in overspread channels, \ac{DD} waveforms are not be well-suited for low-latency applications due to their inherently large temporal occupation. A feasible approach in this case is to shrink the delay axis. However, this compression causes the channel response to wrap over the delay axis, introducing phase ambiguity and tap selectivity due to the quasi-periodicity behavior \cite{OTFS_overspread}. 
\subsection{Birth and Death Phenomena}
In extreme mobility or complex propagation environments, the birth and death phenomenon of channel \ac{MPC}s, where components emerge, overlap, and disappear rapidly, poses a critical challenge for channel modeling \cite{feng2022channel}. This severe time selectivity originating from an ambiguous temporal source, complicates accurate channel characterization and modeling. Estimating and tracking such channels requires short-duration signals and frequent pilot transmissions to counteract the rapid channel aging. Waveforms that restrict their signal spreading within a single symbol duration can accommodate these conditions by adjusting the subcarrier spacing to ensure that the channel remains coherent within each symbol. They can also increase the pilots density per frame to capture rapid variations, such as in \ac{OFDM} and \ac{AFDM}. Conversely, \ac{OTFS} distributes its time domain samples across a block of $N$ symbols, based on the assumption that the number of channel \ac{MPC}s and their delay and Doppler shift values remain constant throughout these symbols. This assumption does not hold in the presence of birth and death phenomena, where different segments of a single \ac{OTFS} symbol in the time domain experience different channel effects. Consequently, the richness and dynamics of such channels are not accurately reflected in the DD domain, since the DD channel response only captures the averaged delay and Doppler shifts over time. This averaging leads to a persistent error floor in \ac{CE}, which is difficult to eliminate or compensate for.
\begin{figure*}[t]
 \centering
  \includegraphics[scale=0.061]{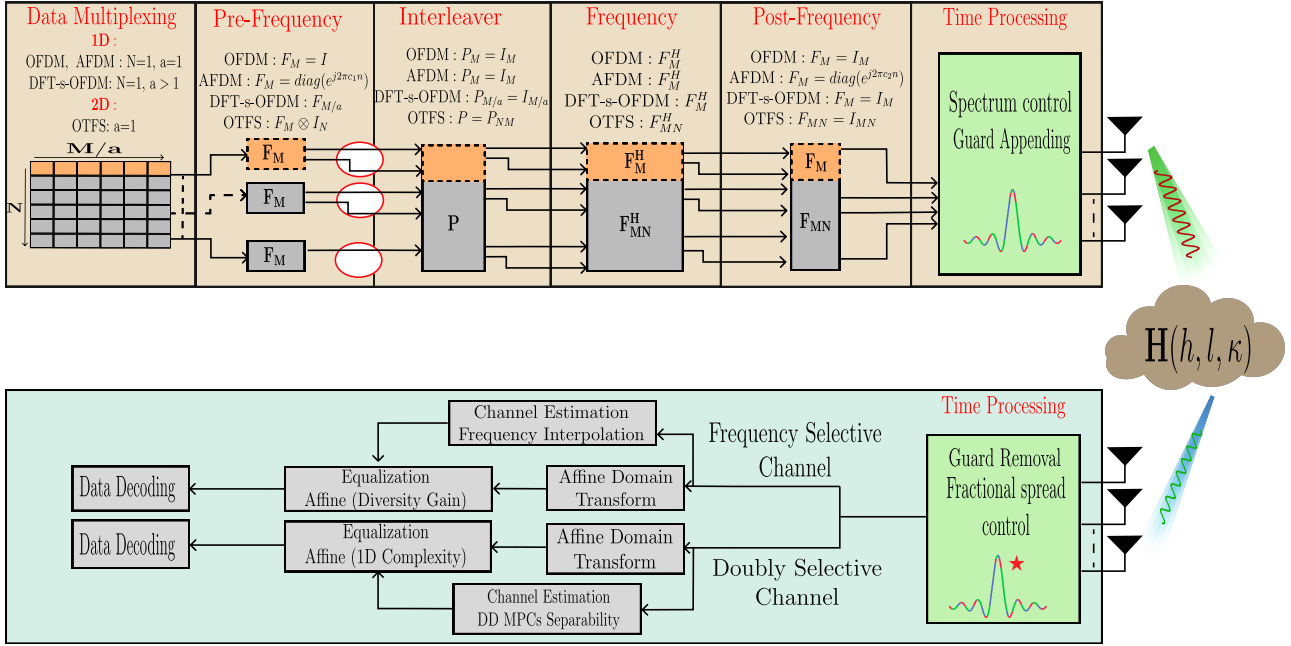}
   \caption{Illustration of the unified waveform generation and the cross-domains communication waveform processing study case : $N,M,\text{ and } P$ denotes symbol spread in time, number of subcarrier per OFDM symbol, and the permutation matrix, respectively. $diag(.)$, $I$ and $a$ denote the diagonal matrix entries, the identity matrix and the DFT-S-OFDM sub-band spread, respectively.}
   \label{fig: gen_model}
\end{figure*}
\section{Unified Waveform Processing}
A compact, unified, and adaptive signal generation framework capable of producing various waveforms including \ac{OFDM}, \ac{OTFS}, and chirp-based variants, has been demonstrated as feasible \cite{solaija2024orthogonal, rou2024orthogonal}. Specifically, by applying pre- and post-spreading operations alongside permutation matrices to the DFT kernel, different spreading mechanisms can be realized, enabling flexible time domain signal shaping. While this approach facilitates waveform unification under the 3GPP-compatible DFT-spreading framework, restricting signal processing to a single domain, from data multiplexing to equalization and symbol detection, proves inefficient due to the divergent signal characteristics required for each operation. Such a constraint limits the full exploitation of channel features available across different domains. Moreover, next-generation use cases such as \ac{ISAC}, low-complexity \ac{IoT} integration, and scalable physical-layer security (PLS) exceed the processing capabilities of conventional OFDM, which is inherently restricted by its \ac{TF} granularity. Consequently, different transform domains can be interchangeably employed for distinct operations, such as data multiplexing, \ac{CE} and equalization, target detection, key generation, and symbol detection, tailored to application-specific requirements and channel conditions. Crucially, this channel-aware processing approach can be seamlessly implemented by exploiting the intrinsic interrelationships among domains, often realized through simple DFT-based transformations, while preserving backward compatibility with 3GPP-standardized waveforms. Furthermore, the localization and spreading properties of signals in alternative domains enable novel channel adaptation techniques, multi-dimensional sensing, and cross-domain key generation schemes. For instance, within the same frequency allocation for \ac{OFDM} signaling, a subset of subcarriers can be populated with a predefined sequence, which, when properly filtered at the receiver, produces an ON-OFF keying (OOK) time domain signal. This permits simultaneous support for heterogeneous users, aligning with recent standardization efforts for wake-up signal (WUS) generation. The unified waveform generation and cross-domain communication processing framework is illustrated in Fig. \ref{fig: gen_model}. 
\section{6G Use Cases}
This section outlines representative use cases for the proposed waveform processing framework. Although the scope of 6G applications extends beyond the scenarios presented herein, the selected cases represent some of the most compelling and practical examples. 
\subsection{Waveform Processing for Communication} 
Conventionally, core operations such as data multiplexing, equalization, and  \ac{CE} are all performed within the same domain. However, under certain channel conditions, single-domain processing is not efficient and limits the advantages of some waveforms. The subsequent section demonstrates a case study for cross-domain \ac{CE} and equalization for \ac{AFDM}, under different channel conditions.

\par
\begin{figure}[t!]
    \centering
    \includegraphics[width=0.45\textwidth]{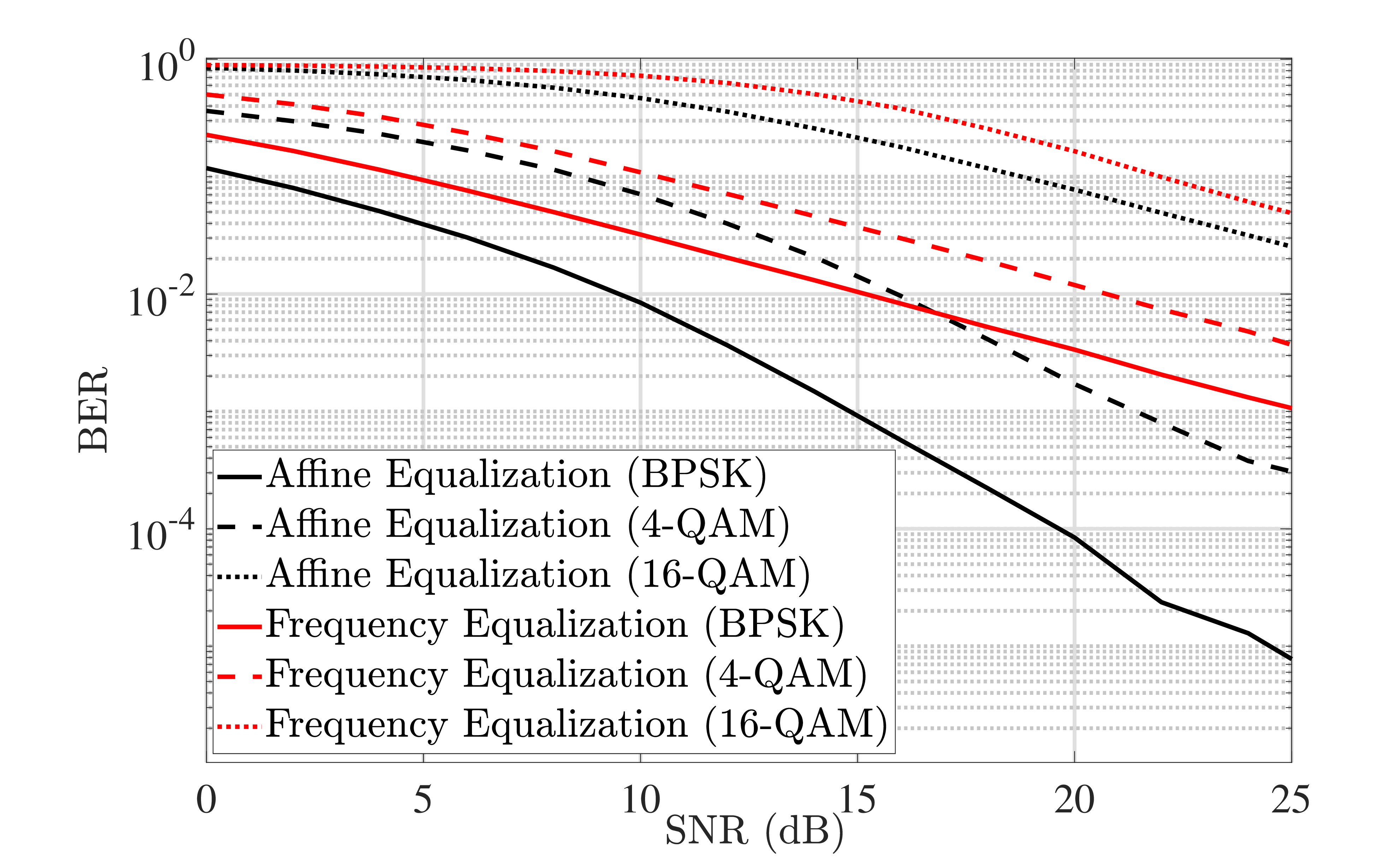}
    \caption{Performance comparison of AFDM and OFDM with frequency domain channel estimation and domain specific equalization over a frequency selective \ac{TDL} 3GPP urban channel model (subcarrier spacing: $\Delta f=30$ KHz, FFT size: $1024$).}
    \label{fig: BER}
\end{figure}

 \textit{\textbf{Static channels}}: Affine domain's \ac{CE} requires an exhaustive search across delay bins, which becomes increasingly complex and latency-sensitive in the presence of high-resolvable MPCs and their fractional components. Alternatively, by performing CE in the frequency domain, all channel delays can be effectively tracked using a simple interpolation. However, the frequency domain offers only first-order diversity and therefore depends heavily on channel coding to ensure reliability. Conversely, full diversity is achieved in the affine domain for static channels, which reduces the dependency on channel coding and thereby enhances achievable throughput. The interplay between the frequency domain CE and the affine domain equalization enables low complexity \ac{CE} while leveraging the diversity benefits of AFDM. Fig. \ref{fig: BER} illustrates diversity gain in the raw \ac{BER} when exploiting affine domain equalization compared to frequency domain equalization while using \ac{CE} in the frequency domain for both cases. The transmitted signal has a block pilot structure, where an all-pilot \ac{AFDM} symbol with known structure in the frequency domain is followed by \ac{AFDM} data symbols.
\par
\begin{figure}[t!]
    \centering
    \includegraphics[width=0.45\textwidth]{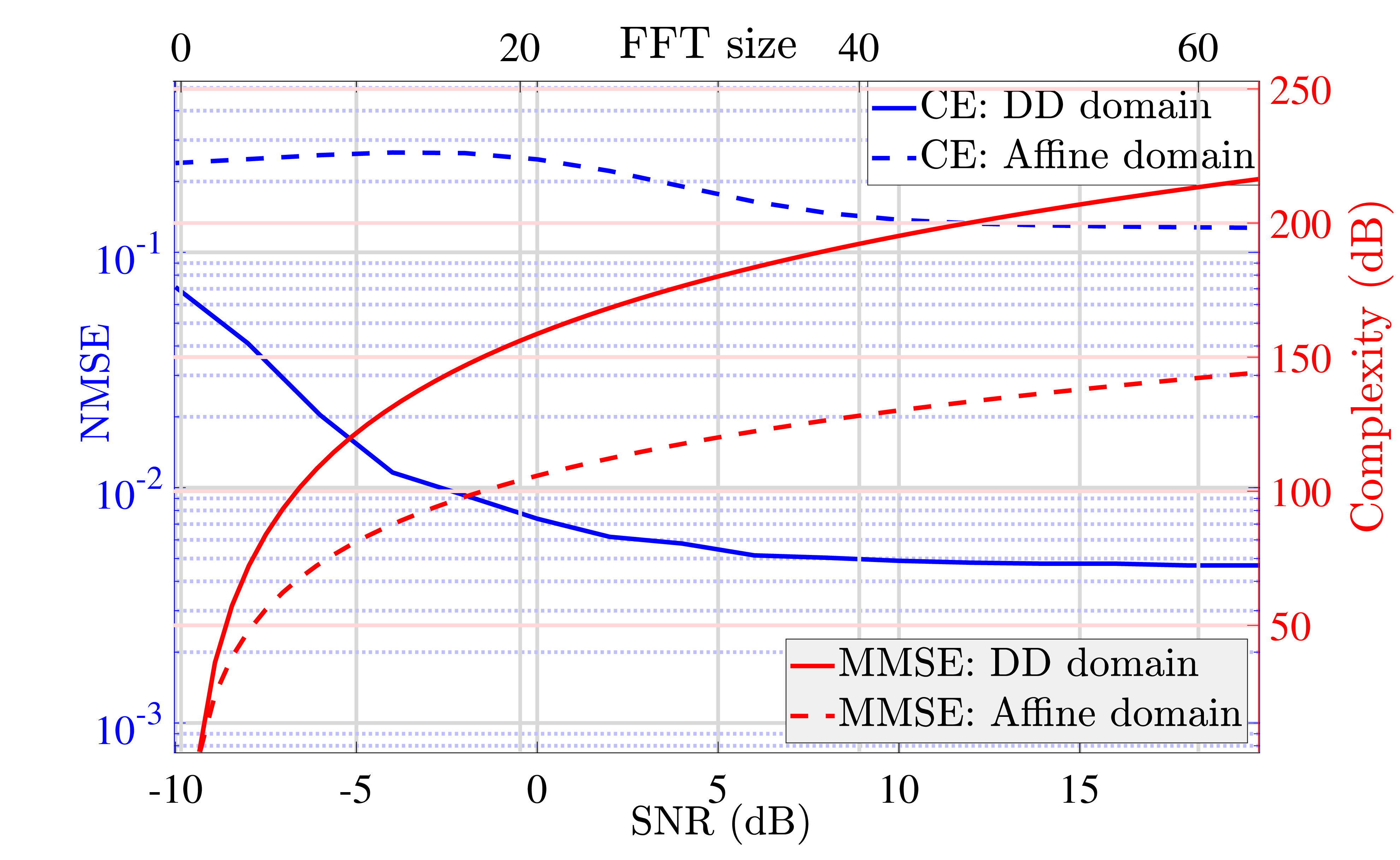}
    \caption{Performance tradeoff between \ac{CE} \ac{NMSE} and equalization complexity between the \ac{DD} domain and the affine domain under the doubly dispersive \ac{EVA} channel model (subcarrier spacing: $\Delta f=30$ KHz, FFT size: $1024$).}
    \label{fig: NMSE}
\end{figure}

 \textit{\textbf{Doubly dispersive channels}}: Due to its inherent 1D structure, \ac{AFDM} encounters ambiguity in accurately estimating the delay and Doppler shift of each channel tap when fractional components are present. Alternatively, the 2D orthogonal structure of \ac{OTFS} facilitates the separation of delay and Doppler shifts. Whereas, even fractional spreads can be effectively managed through proper pulse shaping and sufficient guard intervals, making \ac{CE} in the \ac{DD} domain more reliable compared to the affine domain. However, the 2D structure of OTFS significantly increases the computational complexity and latency of the equalization and detection process. When accurate channel information is available, AFDM achieves OTFS-like \ac{BER} performance, while enabling low-complexity equalization and through its 1D structure. This interchangeability between the two domains enables efficient signal processing in doubly dispersive channels by allowing accurate CE in the \ac{DD} domain along with low-complexity equalization in the affine domain. Fig. \ref{fig: NMSE} shows the superiority of the DD domain in achieving low channel \ac{NMSE} values in the case of a doubly dispersive channel with fractional components compared to the affine domain; simultaneously, \ac{MMSE} equalization in the affine domain ensures a $50$ dB reduction in computational complexity.
\subsection{Waveform Processing for Sensing} 
Sensing in \ac{6G} is expected to operate across a wide range of scenarios, each demanding distinct sensing signal properties. DD-based waveforms offer high Doppler sensitivity due to their temporal spread, enabling the separation of targets ranging from pedestrians to high-speed vehicles. Moreover, chirp-based waveforms exhibit strong time domain correlation properties, allowing low-capable devices to perform sensing without sampling the entire bandwidth. They also support low \ac{SNR} target detection by leveraging time processing gain. Furthermore, CP-\ac{OFDM} has demonstrated superior ranging performance through its ability to achieve very low sidelobes compared to other waveforms \cite{liu2025cp}. Building on these waveform characteristics, a full-CP OTFS frame embedded with a chirp pilot that spreads diagonally in the DD domain can achieve high performance in multi-target sensing, IoT sensing, and object ranging scenarios, especially when processed separately using a DD receiver, a chirp receiver, and an OFDM receiver, respectively. Additionally, integrating sensing signals within the \ac{ISAC} framework introduces an inherent trade-off between communication efficiency and sensing accuracy. For instance, efficient sensing commonly necessitates waveform designs comprising time-localized pulses to enhance range resolution, while employing extended temporal durations to improve Doppler shift estimation accuracy \cite{Fan_pulse}. This extended symbol occupations causes significant latency which is not adequate for \ac{URLLC} applications. Therefore, a waveform design which supports coexistence between longer sensing signal and shorter communication symbols, can be employed through longer chirp signal spanning different consecutive \ac{OFDM} symbols \cite{csahin2020multi}.    

\begin{figure*}[t]
   \centering
   \includegraphics[scale=0.048]{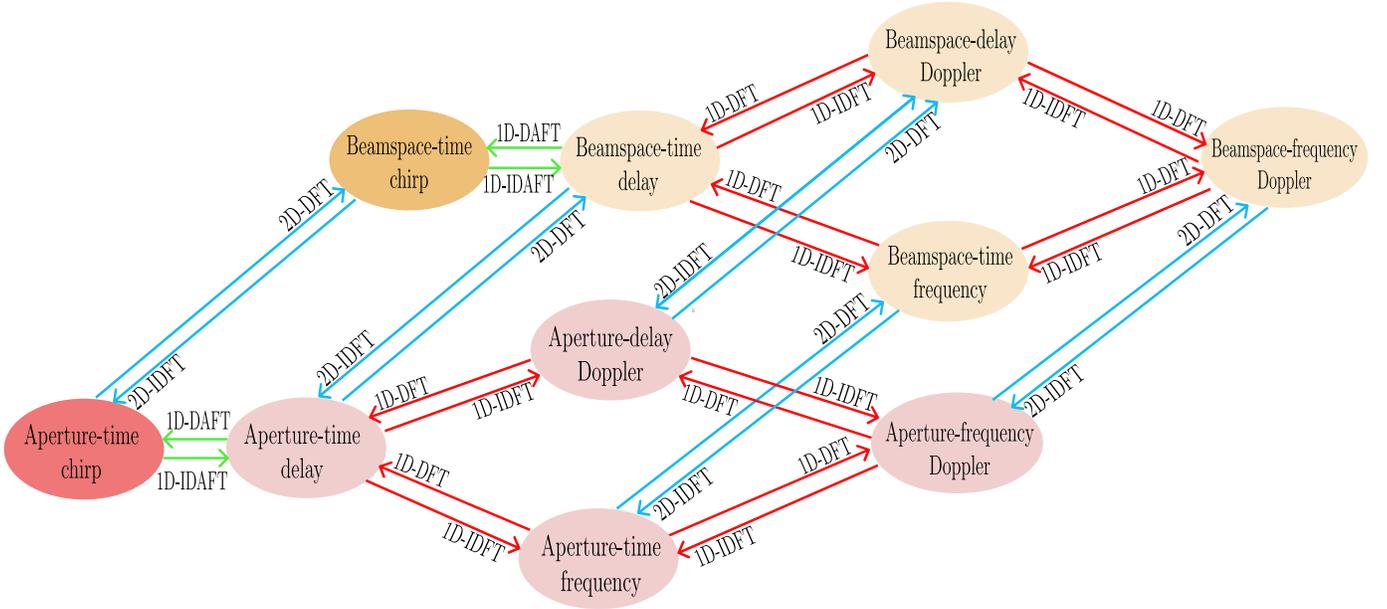}
   \caption{Illustration of the MIMO expansion as a waveform processing domain using DFT operations, and the interchangeable relationship between beamspace, aperture and other physical and logical domains.}
   \label{fig: domian_hhh}
\end{figure*}
\subsection{Multi-dimensional Waveform Processing for MIMO}
The previous subsections described communication and sensing as point-to-point signal transmissions. However, \ac{MIMO} systems are foundational to modern wireless networks and serve as key enablers for waveform performance optimization. \ac{MIMO} extends waveform processing into a three-dimensional space, typically spanning the time, frequency, and spatial domains, by exploiting the spatial characteristics of antenna arrays, represented in either the aperture or the beamspace domain. These two spatial representations are interconnected through 2D \ac{DFT} operations, Fig. \ref{fig: domian_hhh} illustrates that MIMO domain extension is not limited to physical domains but extends also to the logical domains \cite{xiao2024rethinking}. Incorporating MIMO as an additional processing dimension enhances signal handling capabilities for both communication and sensing. In addition to improving throughput, MIMO serves as a dynamic data multiplexer that adapts to channel variations, as it enables the decomposition of a frequency-selective channel into parallel flat-fading subchannels and, under appropriate transformations, allows the Doppler spread to be represented as a single dominant frequency shift. This decomposition reduces the computational burden associated with processing in 2D domains such as the \ac{DD} domain, by enabling efficient parallelization that leverages the inherent sparsity of the DD-beamspace representation. Moreover, MIMO as a processing domain significantly enhances sensing performance. Particularly, the beamspace domain enables angular separation of targets based on their angles of arrival (AOA), thereby mitigating clutter effects and allowing for targets separation using the same physical resources. This capability surpasses the resolution limits imposed by time and frequency domains alone.

\subsection{Waveform Engineering for Security}
While conventional upper-layer encryption and security mechanisms are reliable, they generally introduce additional system complexity and latency. Therefore, the tendency towards adopting PLS has gained momentum lately. PLS exploits the channel and waveform properties to either generate authentication keys or counter malicious attacks. Waveforms with diverse \ac{TF} spreading properties offer inherent robustness against jamming and narrowband attacks, as each data symbol is spread across the entire \ac{TF} plane, where only a portion of the subcarrier is affected. Conversely, in OFDM, as each data subcarrier is limited to a small spectrum portion, narrowband interference is mitigated by repetition coding, significantly degrading the \ac{SE}. PLS also relies on signal entropy to enable the generation of longer and richer authentication keys, used for countering threats such as eavesdropping and spoofing. This entropy can be derived from features embedded in waveform design parameters and effective channel responses. OFDM offers only a first-order diversity, as its  frequency domain effective channel response has limited channel features, where all the MPCs collapse into a single tap. In contrast, alternative 6G waveform candidates enable a more diverse representation of channel parameters, yielding distinctive information for each channel tap. This enhanced parameter diversity facilitates the generation of longer and more unique authentication keys, thereby increasing their resistance to replication and spoofing attempts. Furthermore, the design parameters of \ac{OTFS} and \ac{AFDM} offer an additional DoF which increases the signal entropy. For example, in \ac{OTFS}, the CP structure, whether it is full, reduced, or zero-padded, creates variations in the phases of the effective channel in the DD domain. Similarly, \ac{AFDM} tunes its channel representations via its design parameters, where $c_1$ adjusts the tap spacing and $c_2$ affects the phase variation across the affine domain. Additionally, techniques like directional modulation and noise injection through MIMO beamspace, enable 3 dimensional (3D) signal securing through \ac{TF} spreading and space spots, either by focusing the signal \ac{SNR} or eliminating attackers using noise beams \cite{liu2024artificial}. 
\subsection{IoT Waveform Processing}
Non-coherent detection, asynchronous transmission, and low computational complexity are the defining characteristics of IoT radios that support extended communication range and prolonged battery life. However, these same features hinder seamless integration with existing 3GPP frames, particularly those based on \ac{OFDM}, which typically require coherent reception and moderate processing capabilities. Nevertheless, specific waveform designs can enable the coexistence of IoT devices with cellular networks. For instance, by applying proper bit sequence and frequency shaping at the input of DFT-s-OFDM, an OOK signal overlapped in time with the rest of the OFDM subcarriers can be generated, where IoT devices are able to capture the OOK signal using narrowband filtering. This signal can be detected with either envelope and edge detection or a bank of analog correlators that search for signal's \ac{SNR} maximization. Alternatively, a LoRa-like signal can be generated by changing OCDM carrier indices over multiple symbols sequentially, while the spreading factors can be adjusted by manipulating the chirp-carrier frequency spacing. 

\section{Open Issues and Challenges}
While the unified waveform processing offers a promising approach for the 6G physical layer design, several waveform-related challenges remain open within the challenging 6G requirements.
\begin{itemize}
    \item \textbf{Multiple Access}: Unlike power-domain \ac{NOMA}, which relies on complex successive interference cancellation (SIC) and the acquisition of complete channel state information, waveform-domain \ac{NOMA} has emerged as a promising alternative. This approach enables the superposition of distinct waveforms by leveraging their inherent localization properties, such as chirp signals with diagonal dispersion in the \ac{DD} domain, or \ac{OFDM} signals resembling \ac{AWGN} in the affine domain. Nevertheless, a unified and compact analytical framework for characterizing waveform overlap and orthogonality across different domains remains an open research challenge.

    \item \textbf{Grant-free Access}: To mitigate scheduling overhead and relax strict synchronization constraints, waveform designs that can tolerate marginal timing and frequency offsets while enabling quasi-orthogonal multiple access are essential for supporting massive connectivity and grant-free access. In this context, LoRa chirp-based signaling combined with slotted ALOHA access emerges as a promising solution, although lacking compatibility with the 3GPP physical-layer specifications and suffering from high collision probability.  
    
    \item \textbf{Hardware Complexity}: Limited hardware capability at the user-end (UE) restricts multi-domain signal processing, particularly when considering MIMO and \ac{ISAC} systems. This complexity extends to the receiver filter design, aperture size, and application-based processing blocks.

    \item \textbf{AI-driven Processing}: Cross-domain waveform processing requires frequent channel information acquisition which can lead to significant signaling overhead and latency. Therefore, AI-driven approaches can be employed to track and predict channel conditions, while also enabling adaptive transform-domain tuning.    
\end{itemize}
\section{Conclusion}
A complete transition from OFDM-based physical layer processing in favor of emerging waveform candidates is impractical. Instead, these newly proposed multiplexing domains should be regarded as logical projection spaces that offer enhanced channel representation and unlock capabilities essential for next-generation applications. This work presents a comprehensive analysis of channel characteristics and representations across multiple logical and physical domains, considering both simple and complex propagation environments. Building upon this, we propose an adaptive waveform processing framework that extends the conventional \ac{OFDM} demodulation chain to support different logical domains. The proposed architecture preserves compatibility with existing \ac{OFDM} and DFT-s-OFDM systems through pre- and post-processing matrices. Based on the prevailing channel characteristics and application-specific requirements, this flexible framework enables dynamic processing in appropriate domains. Finally, we identify key use cases for the proposed framework, followed by future challenges and research direction. 

\bibliographystyle{IEEEtran}
\bibliography{Mendeley.bib}

\begin{thebibliography}{10}
\providecommand{\url}[1]{#1}
\csname url@samestyle\endcsname
\providecommand{\newblock}{\relax}
\providecommand{\bibinfo}[2]{#2}
\providecommand{\BIBentrySTDinterwordspacing}{\spaceskip=0pt\relax}
\providecommand{\BIBentryALTinterwordstretchfactor}{4}
\providecommand{\BIBentryALTinterwordspacing}{\spaceskip=\fontdimen2\font plus
\BIBentryALTinterwordstretchfactor\fontdimen3\font minus \fontdimen4\font\relax}
\providecommand{\BIBforeignlanguage}[2]{{%
\expandafter\ifx\csname l@#1\endcsname\relax
\typeout{** WARNING: IEEEtran.bst: No hyphenation pattern has been}%
\typeout{** loaded for the language `#1'. Using the pattern for}%
\typeout{** the default language instead.}%
\else
\language=\csname l@#1\endcsname
\fi
#2}}
\providecommand{\BIBdecl}{\relax}
\BIBdecl

\bibitem{saad2019vision}
W.~Saad, M.~Bennis, and M.~Chen, ``A vision of {6G} wireless systems: Applications, trends, technologies, and open research problems,'' \emph{IEEE network}, vol.~34, no.~3, pp. 134--142, 2019.

\bibitem{3gpp_tr_38_900}
\BIBentryALTinterwordspacing
``{LTE; 5G; Study on channel model for frequency spectrum above 6 GHz (3GPP TR 38.900 version 14.2.0 Release 14)},'' {3rd Generation Partnership Project (3GPP)}, Tech. Rep. TR 38.900, 2017, version 14.2.0.
\BIBentrySTDinterwordspacing

\bibitem{solaija2024orthogonal}
M.~S.~J. Solaija, S.~E. Zegrar, and H.~Arslan, ``Orthogonal frequency division multiplexing: The way forward for {6G} physical layer design?'' \emph{IEEE Vehicular Technology Magazine}, 2024.

\bibitem{xiao2024rethinking}
Z.~Xiao, X.~Liu, Y.~Zeng, J.~A. Zhang, S.~Jin, and R.~Zhang, ``Rethinking waveform for {6G}: Harnessing delay-doppler alignment modulation,'' \emph{IEEE Communications Magazine}, 2024.

\bibitem{deng2025unifying}
Q.~Deng, Y.~Ge, and Z.~Ding, ``A unifying view of {OTFS} and its many variants,'' \emph{IEEE Communications Surveys \& Tutorials}, 2025.

\bibitem{AFDM_fractional}
A.~Bemani, N.~Ksairi, and M.~Kountouris, ``Integrated sensing and communications with affine frequency division multiplexing,'' \emph{IEEE Wireless Communications Letters}, 2024.

\bibitem{arous2024novel}
A.~Arous, H.~Haif, and H.~Arslan, ``A novel channel estimation for chirp-based multicarrier waveforms under practical pulse shaping,'' \emph{Authorea Preprints}, 2024.

\bibitem{OTFS_fraction}
S.~E. Zegrar, A.~S. S{\"u}mer, and H.~Arslan, ``Fractional delay \& fractional doppler estimation and mitigation framework in {OTFS} systems,'' \emph{IEEE Transactions on Vehicular Technology}, 2024.

\bibitem{OTFS_overspread}
P.~Priya, Y.~Hong, and E.~Viterbo, ``{OTFS} channel estimation and detection for channels with very large delay spread,'' \emph{IEEE Transactions on Wireless Communications}, 2024.

\bibitem{feng2022channel}
Y.~Feng, N.~Ge, X.~Tao, Q.~Song, and T.~Xiang, ``Channel modelling for {v2v} highway scenario based on birth and death process,'' \emph{Wireless Communications and Mobile Computing}, vol. 2022, no.~1, p. 3384362, 2022.

\bibitem{rou2024orthogonal}
H.~S. Rou, G.~T.~F. de~Abreu, J.~Choi, D.~González~G., M.~Kountouris, Y.~L. Guan, and O.~Gonsa, ``From orthogonal time–frequency space to affine frequency-division multiplexing: A comparative study of next-generation waveforms for integrated sensing and communications in doubly dispersive channels,'' \emph{IEEE Signal Processing Magazine}, vol.~41, no.~5, pp. 71--86, 2024.

\bibitem{liu2025cp}
F.~Liu, Y.~Zhang, Y.~Xiong, S.~Li, W.~Yuan, F.~Gao, S.~Jin, and G.~Caire, ``{CP-OFDM} achieves the lowest average ranging sidelobe under {QAM/PSK} constellations,'' \emph{IEEE Transactions on Information Theory}, 2025.

\bibitem{Fan_pulse}
Z.~Liao, F.~Liu, S.~Li, Y.~Xiong, W.~Yuan, C.~Masouros, and M.~Lops, ``Pulse shaping for random {ISAC} signals: The ambiguity function between symbols matters,'' \emph{IEEE Transactions on Wireless Communications}, 2025.

\bibitem{csahin2020multi}
M.~M. {\c{S}}ahin and H.~Arslan, ``Multi-functional coexistence of radar-sensing and communication waveforms,'' in \emph{2020 IEEE 92nd Vehicular Technology Conference (VTC2020-Fall)}.\hskip 1em plus 0.5em minus 0.4em\relax IEEE, 2020, pp. 1--5.

\bibitem{liu2024artificial}
K.~Liu, J.~Chen, and X.~Lei, ``Artificial noise aided directional modulation: A transmitter architecture perspective,'' \emph{IEEE Transactions on Communications}, 2024.

\end{thebibliography}

\end{document}